\documentclass[A4,12pt]{article}
\usepackage{graphicx}
\usepackage{rotating}
\usepackage{amssymb}
\usepackage{mathptmx}
\usepackage[numbers]{natbib}
\usepackage[nofighead,nomarkers]{endfloat}
\newcommand{\E}{\mathcal{E}}

\bibpunct{}{}{,}{s}{}{,}

\begin{document}

\title{ New version of quantum mechanics at
finite temperatures as a ground for  description of nearly perfect
fluids\\}

\author{A.D. Sukhanov$^1$ \and O.N. Golubjeva $^2$    }


\date{15.06.2010}

\maketitle


\begin{abstract}

We  suggest  a more general than quantum statistical mechanics
($QSM$) microdescription of objects in a heat bath taken into
account a vacuum as an object environment - modification of quantum
mechanics at finite temperatures; we call it $(\hbar, k)$-dynamics
($ \hbar kD$). This approach allows  us in a new manner   to
calculate some important macroparameters and to   modify standard
thermodynamics. We create an effective apparatus for   features
description  of nearly perfect fluids  in various mediums. As an
essentially new model of an object environment we suppose a quantum
heat bath and its properties, including cases of cold and warm
vacuums,  are studied. We   describe the   thermal equilibrium state
in place of the traditional  density operator in term of  a wave
function the amplitude and phase of which  have  temperature
dependence.

We introduce a new generative operator, Schroedingerian,  or
stochastic action operator, and show its fundamental role in the
microdescription.   We  demonstrate that a new  macroparameter,
namely the effective action , can be obtained through averaging  of
the Schroedingerian  over the   temperature dependent wave function.
It is established that such different parameters as internal energy,
effective temperature, and effective entropy and their fluctuations
can be expressed through a single  quantity - the effective action.

\end{abstract}

\section{Fundamental microtheory: ($\hbar, k$)-
dynamics} We develop a version of the universal theory that is
unlike to the particular   theory proposed earlier  for  explaining
of ratio "shear viscosity to specific entropy' in quark-gluon
plasma.    We propose to formulate a quantum-thermal dynamics or,
briefly, ($\hbar,k$)- dynamics ($\hbar kD$), as a modification of
standard quantum mechanics taking thermal effects into account. The
principal distinction of such a theory from QSM is that in it  the
state of a microobject under the conditions of contact with the
quantum heat bath (QHB) is generally described not by the density
matrix but by a temperature-dependent complex wave function.

We note that this is not a "thechnical sleight-of-hand". Using  the
wave function, we thereby suppose to consider pure and mixt states
simultaneously in the frame of Gibbs' ensemble. It is in principle
differs from  Boltzmann' assembly used in QSM.

A general idea of our investigation: to construct a theory it is
necessary

 1. to change $\hat\rho(T) \Rightarrow \Psi_T(q);$

2. to introduce (except of Hamiltonian) a new operator - the
stochastic action operator $\hat j;$

3. to use an idea of  heat bath at $T=0$ ("cold"  heat bath) also;

4. to use an idea of vacuum at $T>0$ ("thermal" vacuum) also.

This theory is based on a new microparameter, namely, the stochastic
action operator. In this case, we demonstrate that averaging the
corresponding microparameters over the temperature-dependent wave
function, we can find the most important effective macroparameters,
including internal energy, temperature, and entropy. They have the
physical meaning of the standard thermodynamic quantities using in
the phenomenological macrodescription.

\subsection {{The model of the QHB: a case of the "cold"
vacuum}}

To describe the environment with the holistic stochastic action that
was previously called the thermal field vacuum by Umezawa, we
introduce a concrete model, the QHB.   According to this, the QHB is
a set of weakly coupled quantum oscillators with all possible
frequencies. The equilibrium thermal radiation can serve as a
preimage of such a model in nature.

 The specific feature of our understanding of this model is that we assume that we must apply it
to both the "thermal" ($T\neq  0$) and the "cold" ($T = 0$) vacua.
Thus, in the sense of Einstein, we proceed from a more general
understanding of the thermal equilibrium, which can, in principle,
be established for any type of environmental stochastic action
(purely quantum, quantum-thermal, and purely thermal).

We begin our presentation by studying the "cold" vacuum and
discussing the description of a single quantum oscillator from the
number of oscillators forming the QHB model for $T= 0$ from a new
standpoint.

But we recall that the lowest state    in the energetic
($\Psi_n(q)$) and   coherent states (CS) is the same. In the
occupation number representation, the "cold" vacuum in which the
number of particles is $n=0$ corresponds to this state. In the $q$
representation, the same ground state of the quantum oscillator is
in turn described by the real wave function

\begin{equation}\label{4}
\Psi_0(q) =[2\pi (\Delta q_0)^2]^{-1/4}\exp \left\{-\frac{q^2}
{4(\Delta q_0)^2}\right\}.
\end{equation}
As is well known, CS are the eigenstates of the non-Hermitian
particle annihilation operator $\hat a$ with complex eigenvalues.
But they include one isolated state $|0_a\rangle =
|\Psi_0(q)\rangle$ of the particle vacuum in which eigenvalue of
$\hat a$  is   zero

\begin{equation}\label{5}
\hat{a}|0_a\rangle=0|0_a\rangle;\;\;\;\;\;\hat{a}|\Psi_0(q)\rangle
=0|\Psi_0(q)\rangle.
\end{equation}

In what follows, it is convenient to describe the QHB in the $q$
representation. Therefore, we express the annihilation operator
$\hat a$ and the creation operator $\hat a^\dagger$  in terms of the
operators $\hat p$   and $\hat q$  using the traditional method. We
have

\begin{equation}\label{6}
\hat a=\frac1 2 \left (\frac{\hat p}{\sqrt{\Delta
p_0^2}}-i\frac{\hat q}{\sqrt{\Delta q_0^2}}\right);\;\;\;\hat
a^\dag=\frac1 2 \left (\frac{\hat p}{\sqrt{\Delta
p_0^2}}+i\frac{\hat q}{\sqrt{\Delta q_0^2}}\right).
\end{equation}
The particle number operator then becomes
\begin{equation}\label{7} \hat N_a=\hat a^\dag\hat
a=\left(\frac{\hat p^2}{\Delta p_0^2} -\frac 12\hat I +\frac{\hat
q^2}{\Delta q_0^2}\right)=\frac{1}{\hbar\omega}\left(\frac{\hat
p^2}{2m}-\frac{\hbar\omega}{2}\hat I + \frac{m\omega^2\hat
q^2}{2}\right).
\end{equation}

The sum of the first and third terms in the parentheses forms the
Hamiltonian $\mathcal{H}$  of the quantum oscillator, and after
multiplying relation (7) by $\hbar\omega$  on the left and on the
right, we obtain the standard interrelation between the expressions
for the Hamiltonian in the $q-$ and $n-$ representations:

\begin{equation}\label{8} \hat{\mathcal{H}}=\frac{\hat
p\,^2}{2m}+\frac{m\omega^2\hat q^2}{2}=\hbar\omega(\hat
N_a+\frac12\hat I),
\end{equation}
where $\hat I$  is the unit operator.

From the thermodynamics standpoint, we are concerned with the
internal energy of the quantum oscillator in equilibrium with the
"cold" QHB. Its value is equal to the mean of the Hamiltonian
calculated over the state $|0_a\rangle\equiv|\Psi_0(q)\rangle$ :

\begin{equation}\label{9}
U_0=\langle\Psi_0(q)|\hat{\mathcal H}|\Psi_0(q)\rangle=\hbar\omega
\langle\Psi_0(q)|\hat
{N}_a|\Psi_0(q)\rangle+\frac{\hbar\omega}{2}=\frac{\hbar\omega}{2}
=\varepsilon_0.
\end{equation}

It follows from formula (9) that in the given case, the state
without particles coincides with the state of the Hamiltonian with
the minimum energy $\varepsilon_0$. The quantity $\varepsilon_0$,
traditionally treated as the energy of zero oscillations, takes the
physical meaning of the internal energy $U_0$  of the quantum
oscillator in equilibrium with the "cold" vacuum.

\subsection { {  Passage to the "thermal" vacuum}} We can pass
from the "cold"  to the "thermal"  vacuum in the spirit of Umezawa
using the Bogoliubov $(u, v)-$ transformation with the
temperature-dependent coefficients

\begin{equation}\label{11}
u=\left(\frac12\coth\frac{\hbar\omega}{2k_BT}+\frac12\right)^{1/2}e^{\textstyle
{i\frac{\pi}{4}}};\;\;\;\;v=\left(\frac12\coth\frac{\hbar\omega}{2k_BT}-\frac12\right)^{1/2}e^{-\textstyle
{i\frac{\pi}{4}}}.
\end{equation}

In the given case, this transformation is canonical but leads to a
unitarily nonequivalent representation because the QHB at any
temperature is a system with an infinitely   number of freedom
degrees.

In the end, such a transformation reduces to passing from the set of
quantum oscillator CS to a more general set of states called the
thermal correlated coherent states  (TCCS). They are selected
because they ensure that the Schrodinger coordinate-momentum
uncertainty relation is saturated at any temperature. From the of
the second-quantization apparatus standpoint, the Bogoliubov $(u,
v)-$ transformation ensures the passage from the original system of
particles with the "cold" vacuum  $|0_a\rangle$ to the system of
quasiparticles described by the annihilation operator $\hat b$ and
the creation operator $\hat b^{\dag}$ with the "thermal" vacuum
$|0_b\rangle$.

To obtain from "cold"  vacuum to "thermal"  one using $(u,v)$ -
Bogolubov's transformations it is necessary to pass:

1. from CS to TCCS: $\Psi_0(q)\Rightarrow\Psi_T(q); \;\;\;\;
|0_a\rangle\Rightarrow||0_b\rangle; $ 2. from particles to
quasiparticles: $\hat a \Rightarrow \hat b = \hat b(T). $

In this case, the choice of transformation coefficients (10) is
fixed by the requirement that for any method of description, the
expression for the mean energy of the quantum oscillator in thermal
equilibrium be defined by the Planck formula, which can be obtained
from experiments:

\begin{equation}\label{11}
\E_{Pl.}=\hbar\omega(\exp{\frac
{\hbar\omega}{k_BT}}-1)^{-1}+\frac{\hbar\omega}{2}
=\frac{\hbar\omega}{2}\coth\frac{\hbar\omega}{2k_BT}.
\end{equation}

Earlier was shown by us, the state of the "thermal" vacuum
$|0_b\rangle \equiv|\Psi_{\scriptscriptstyle{T}}(q)\rangle$ in the
$q-$ representation corresponds to the complex wave function

\begin{equation}\label{12}
\Psi_{\scriptscriptstyle{T}}(q)=[2\pi (\Delta q)^2]^{-1/4}\exp
\left\{-\frac{q^2} {4(\Delta q)^2}(1-i\alpha)\right\},
\end{equation}
where

\begin{equation}\label{13}
(\Delta q)^2 = \frac{\hbar}{2m\omega} \coth\frac
{\hbar\omega}{2k_BT};\;\;\;\;\alpha=\left[\sinh\frac{\hbar\omega}{2k_BT}\right]^{-1}
;\;\;\;\; (\Delta p)^2 = \frac{\hbar m\omega}{2} \coth\frac
{\hbar\omega}{2k_BT}.
\end{equation}

We note that the expressions for the probability densities
$\rho_{\scriptscriptstyle{T}}(q)$ and
$\rho_{\scriptscriptstyle{T}}(p)$ have already been obtained by
Bloch, but the expressions for the phase that depend on the
parameter $\alpha$ play a very significant role and were not
previously known. It is also easy to see that as $T \rightarrow 0$,
the parameter  $\alpha \rightarrow 0$ and the function
$\Psi_{\scriptscriptstyle{T}}(q)$  from the set of TCCS passes to
the function $\Psi_{\scriptscriptstyle{0}}(q)$  from the set of CS.

Of course, the states from the set of TCCS are the eigenstates of
the non-Hermitian quasiparticle annihilation operator  $\hat b$ with
complex eigenvalues. They also include one isolated state of the
quasiparticle vacuum  in which eigenvalue of $\hat b$ is zero,

\begin{equation}\label{15}
\hat{b}|0_b\rangle=0|0_b\rangle; \;\;\;\;\hat{b}|\Psi_T(q)\rangle
=0|\Psi_T(q)\rangle.
\end{equation}

Using condition (15) and expression (12) for the wave function of
the "thermal"  vacuum, we obtain the expression for the operator
$\hat b$ in the $q-$ representation:

\begin{equation}\label{16}
\hat b=\frac 12\sqrt{\coth
\frac{\hbar\omega}{2k_BT}}\left[\frac{\hat p}{\sqrt{\Delta
p_0^2}}-i\frac{\hat q}{\sqrt{\Delta q_0^2}} (\coth
\frac{\hbar\omega}{2k_BT})^{-1}(1-i\alpha)\right].
\end{equation}
The corresponding quasiparticle creation operator has the form
\begin{equation}\label{17}
\hat b^{\dag}=\frac {1}{2}\sqrt{\coth
\frac{\hbar\omega}{2k_BT}}\left[\frac{\hat p}{\sqrt{\Delta
p_0^2}}+i\frac{\hat q}{\sqrt{\Delta q_0^2}} (\coth
\frac{\hbar\omega}{2k_BT})^{-1}(1+i\alpha)\right].
\end{equation}

We can verify that as $T\rightarrow 0$ , the operators $\hat
b^{\dag}$ and $\hat b$ for quasiparticles pass to the operators
$a^{\dag}$ and $\hat a$ for particles and $|0_b\rangle\Rightarrow
|0_a\rangle;\;\;\;\;\Psi_T(q)\Rightarrow\Psi_0(q).$

Acting just as above, we obtain the expression for the quasiparticle
number operator in the $q-$ representation

\begin{equation}\label{18}
\hat N_b=\hat b^{\dag}\hat b=\frac 14\coth
\frac{\hbar\omega}{2k_BT}\left[\frac {\hat p^2}{\Delta
p_0^2}-2(\coth \frac{\hbar\omega}{2k_BT})^{-1} (\hat I+\frac \alpha
\hbar\{\hat p,\hat q\})+\frac {\hat q^2}{\Delta q_0^2}\right],
\end{equation}
where we take  $1+\alpha^2=\coth^2\frac{\hbar\omega}{2k_BT}$ into
account when calculating the last term.

\subsection {{ Hamiltonian in TCCS}}

Passing from the quasiparticle number operator to the original
Hamiltonian and multiplying by $\hbar\omega$, we obtain

\begin{equation}\label{19}
\hat{\mathcal{H}}=\hbar\omega(\coth
\frac{\hbar\omega}{2k_BT})^{-1}\left[\hat N_b+\frac 12 (\hat I+
\frac\alpha \hbar\{\hat p,\hat q\})\right].
\end{equation}

We stress that the operator $\{\hat p,\hat q\}$ in formula (19) can
also be expressed in terms of bilinear combinations of the operators
$\hat b^{\dag}$ and $\hat b$, but they differ from the quasiparticle
number operator $N_b$. This means that the operators $\hat{\mathcal
H}$ and $\hat N_b$ do not commute and that the wave function of form
(12) characterizing the state of the "thermal" vacuum is therefore
not the eigenfunction of the Hamiltonian.

As before, we are interested in the thermodynamic quantity, namely,
the internal energy $U$ of the quantum oscillator now in thermal
equilibrium with the "thermal"  QHB. Calculating it just as earlier,
we obtain

\begin{equation}\label{20}
U= \hbar\omega(\coth
\frac{\hbar\omega}{2k_BT})^{-1}\left[\langle\Psi_{\scriptstyle
T}(q)|\hat N_b|\Psi_{T}(q)\rangle+\frac{1}{2} +\frac{\alpha}{2\hbar}
\langle\Psi_{\scriptstyle T}(q)|\{\hat p,\hat q\}
|\Psi_{T}(q)\rangle\right]
\end{equation}
in the $q-$ representation. Because we averaging over the
quasiparticle vacuum in formula (20), the first term in it vanishes.
At the same time, it was shown earlier by us  that

$ \langle\Psi_{\scriptstyle T}(q)|\{\hat p,\hat
q\}|\Psi_{T}(q)\rangle=\hbar\alpha. $

As a result, we obtain the expression for the internal energy of the
quantum oscillator in the "thermal" QHB in the $\hbar kD$:

\begin{equation}\label{22} U=\frac {\hbar\omega}{2 (\coth
\frac{\hbar\omega}{2k_BT})}(1+\alpha^2) =\frac{\hbar\omega}{2}\coth
\frac{\hbar\omega}{2k_BT}=\E_{Pl.},
\end{equation}
where $\E_{Pl.}$ is defined by Planck formula (11). This means that
the average energy of the quantum oscillator at $T\neq 0$ has the
thermodynamic meaning of its internal energy in the case of
equilibrium with the "thermal" QHB. As $T\rightarrow 0$, it passes
to a similar quantity corresponding to equilibrium with the "cold"
QHB.

\section{ {New fundamental operator - Schroedingerian }}

\subsection {{ Schroedinger uncertainties relation}}
Because the original statement of the $\hbar kD$ is the idea of the
holistic stochastic action of the QHB on the object, we introduce a
new operator in the Hilbert space of microstates to implement it.

We recall the general expression of Schwartz inequality
$
|A|^2\cdot|B|^2\geqslant |A\cdot B|^2.
$

Saturated Schroedinger uncertainties relations (SUR)
coordinate-momentum following from it is:

\begin{equation}\label {23}
(\Delta p)^2(\Delta q)^2 = |\tilde
R_{qp}|^2\equiv\sigma^2+\frac{\hbar^2}{4}.
\end{equation}

In the absent of stochastic action $\tilde{R}_{qp}\equiv0.$ As
leading considerations, hereinafter  we use an analysis of the
right-hand side of the  saturated SUR coordinate-momentum.

\subsection{{The stochastic action operator
(Schroedingerian)}}
For not only a quantum oscillator in a QHB but
also any object, the complex quantity in the right-hand side of (23)

\begin{equation}\label{24}
\widetilde{R}_{p\,q}=\langle\Delta p|\Delta q\rangle\qquad\mbox {or
}\widetilde{R}_{p\,q}=\langle\,|\Delta \widehat{p}\,\Delta
\widehat{q}\,|\,\rangle
\end{equation}
has a double meaning. On one hand, it is the amplitude of the
transition from the state $|\Delta q\rangle $ to the state $|\Delta
p\rangle $; on the other hand, it can be treated as   the
Schroedinger quantum correlator calculated over an arbitrary state
$|\; \rangle$  of some operator.

As is well known, the nonzero value of quantity (24) is the
fundamental attribute of nonclassical theory in which the
environmental stochastic action on an object plays a significant
role. Therefore, it is quite natural to assume that the averaged
operator in formula (24) has a fundamental meaning. In view of
dimensional considerations, we call it the stochastic action
operator or Schroedingerian,

\begin{equation}\label{25}
\widehat{j}\equiv\Delta \widehat p\,\Delta \widehat{q}.
\end{equation}

Of course, it should be remembered that the operators $\Delta \hat q
$ and $\Delta\hat p $ do not commute and their product is a
non-Hermitian operator.

To analyze further, following Schroedinger, we can write the given
operator

\begin{equation}\label{26}
\widehat{j}= \frac{1}{2} \left \{\Delta\widehat{p}\Delta\widehat{q}+
\Delta\widehat{q}\Delta\widehat{p}\,\right\}
+\frac{1}{2}\left[\Delta\widehat{p}\Delta\widehat{q}-
\Delta\widehat{q}\Delta\widehat{p}\,\right]=
\widehat{\sigma}-i\,\widehat{j}_{\scriptscriptstyle 0}.
\end{equation}
It allows separating the Hermitian part of $\hat j$ from the
anti-Hermitian one. Then the Hermitian operators $\widehat\sigma$
and $\hat j_0$ have the form

\begin{equation}\label{27}
\widehat{\sigma}\equiv \frac12\{\Delta \hat p,\;\Delta \hat q\};
\;\;\;\;\hat{j}_{\scriptscriptstyle 0} \equiv \frac{ i}{2}\;
[\hat{p}\,,\hat{q}\,]\, \,\ = \frac{\hbar}{2}\;\hat {I}.
\end{equation}
It is easy to see that the mean
$\sigma=\langle\,|\widehat{\sigma}|\,\rangle$ of the operator
$\hat\sigma$ resembles the expression for the standard correlator of
coordinate and momentum fluctuations in classical probability
theory. It transforms into this expression if the operators $\Delta
\hat q $ and $\Delta \hat p $   are replaced with $c$-numbers. It
reflects the contribution to the transition amplitude $\tilde
R_{pq}$ of the environmental stochastic action. Therefore, we call
the operator $\hat{\sigma}$ the external action operator in what
follows.

At the same time, the operators $\hat j_0$ and $\hat j $ were not
previously introduced. The operator $\hat j_0$ of form (27) reflects
a specific peculiarity of the objects to be "sensitive"  to the
minimum stochastic action of the "cold"  vacuum and to respond to it
adequately regardless of their states. Therefore, it should be
treated as a minimum stochastic action operator. Its mean $J_0 =
\frac {\hbar}{2}$ is independent of the choice of the state over
which the averaging is performed, and it hence has the meaning of
the invariant eigenvalue of the operator $\hat j_0$.

\section{ {Effective action as a fundamental generative
macroparameter} }

\subsection{{The mean of the operator $\widehat j$ }}
We now  construct the macrodescription of objects using their
microdescription in the $\hbar kD$. It is easy to see that the mean
$\tilde J$ of the operator $\hat j$ of form (26) coincides with the
complex transition amplitude $\widetilde R_{pq}$ or Schroedinger's
correlator and, in thermal equilibrium, can be expressed as

\begin{equation}\label{28}
\tilde J=\langle \Psi_{\scriptstyle
T}(q)|\,\widehat{j}\,|\Psi_{\scriptstyle T}(q)\rangle=\sigma -
\,i\,J_0,
\end{equation}
where $\sigma $ and $J_0$  are the means of the corresponding
operators. In what follows, we regard the modulus of the complex
quantity  $\widetilde J$,

\begin{equation}\label{29}
|\tilde
J|=\sqrt{\sigma^2+J_0^2}=\sqrt{\sigma^2+\frac{\hbar^2}{4}}\equiv
J_{ef.}
\end{equation}
as a new macroparameter and call it the effective action. It has the
form

\begin{equation}\label{30}
J_{ef.}=\frac\hbar
2\coth\frac{\hbar\omega}{2k_BT}=\frac{U}{\omega}=\frac{k_BT_{ef}}{\omega}
\end{equation}
for the quantum oscillator and coincides with a similar quantity
previously postulated from intuitive considerations.

\subsection{{ Effective entropy in the  $\hbar kD$ }}

The possibility of introducing entropy in the $\hbar kD$ is also
based on using the wave function instead of the density operator.
Using the dimensionless expressions for $\rho (q)= |\Psi(q)|^2$  and
$\rho(p) = |\Psi(p)|^2$, we propose to define a formal coordinate -
momentum entropy $S_{qp}$ by the equality

\begin{equation}\label{41}
S_{qp}\,=-k_B\left\{\int\tilde{\rho}(\tilde{q})
\ln\tilde{\rho}(\tilde{q})
d\tilde{q}+\int\tilde{\rho}(\tilde{p})\ln\tilde{\rho}(\tilde{p})
d\tilde{p}\right\}.
\end{equation}
Substituting the corresponding expressions for $\tilde{\rho} (\tilde
q) $  and  $\tilde{\rho}(\tilde p)$  in (41), we obtain

\begin{equation}\label{42}
S_{qp}=k_B \left\{(1+\ln\frac{2\pi}{\delta})
+\ln\coth\frac{\hbar\omega}{2k_BT}\right\}.
\end{equation}
Obviously, the final result depends on the choice of the constant
$\delta$.

Choosing $\delta = 2\pi$, we can interpret expression (42) as the
quantum-thermal entropy or, briefly, the $QT-$  entropy $S_{QT}$
because it coincides exactly with the effective entropy $S_{ef.}$
obtained earlier by us in the macrotheory framework.

\begin{equation}\label{43}
S_{QT}\equiv
S_{ef.}=k_B\left\{1+\ln\frac{J_{ef.}}{J_0}\right\}=k_B\left\{1+\ln\Omega\right\},
\end{equation}
where $\Omega$ according to Boltzmann is a number of microstates in
the given macrostate. This ensures the consistency between the main
results of our proposed micro- and macrodescriptions and their
correspondence to experiments.

\subsection{ { Quantum Statistical Thermodynamics on the
base of the effective action}}

The above presentation shows that using the $\hbar kD$ developed
here, we can introduce the effective action $J_{ef.} $ as a new
fundamental macroparameter. The advantage of  this quantity is that
it has a microscopic preimage, namely, the stochastic action
operator $\hat j$, or Schroedingerian, which has an obvious physical
sense. Moreover, we can in principle express the main thermodynamic
characteristics of objects in thermal equilibrium in terms of it. As
is well known, temperature and entropy are the most fundamental of
them.

If the notion of effective action is used, these heuristic
considerations can acquire an obvious meaning. For this, we turn to
expression (34) for $T_{ef.}\sim J_{ef.}$. It follows from it that
the effective action is also an \emph{intensive}  macroparameter
characterizing the stochastic action of the "thermal" QHB.

In view of this, the Zero Law of equilibrium quantum Statistical
Thermodynamics can be rewritten as

\begin{equation}\label{48}
J_{ef.}= J_{ef.}^{therm.} \pm \Delta J_{ef.},
\end{equation}
where  $J^{term.}_{ef.}$ is effective action of  QHB, $J_{ef.}$ and
$\Delta J_{ef.}$ are the means of the effective action of an object
and the standard deviation from it. The state of thermal equilibrium
can actually be described in the sense of Newton, assuming that "the
stochastic action is equal to the stochastic counteraction" in such
cases.

\section{ { Connection with experiment}}

Nowadays there is alot of papers on the theory of nearly perfect
fluids where the ratio of shift viscosity  to entropy volume density
is the subject of interest. It was shown by us that this quantity
can be given by
\begin{equation}\label{51}
\frac{J^{ef.}}{S^{ef.}}=\frac{J^{ef.}_{min}}{S^{ef.}_{min}}\;\cdot\frac{\coth
T^{ef.}_{min}/ T}{1+\ln \coth T^{ef.}_{min}/ T}=
\varkappa\;\cdot\frac{\coth\varkappa\omega/T}{1+\ln\coth
\varkappa\omega/T}\rightarrow \varkappa.
\end{equation}

In this expression,

\begin{equation}\label{52}
\varkappa\equiv\frac{J^{ef.}_{min}}{S^{ef.}_{min}}=\frac{\hbar}{2k_B}=3,82\cdot
10^{-12}  \mbox{K c} .
\end{equation}
is the limiting ratio for  $T\ll T^{ef.}. $

In our opinion, the quality is not only the notation for one of the
possible combinations of the world constants $\hbar$ and $k_B$. It
also has its intrinsic physical meaning. We sure that the quantity
$\varkappa$ plays the role of a constant essentially characterizing
the holistic stochastic action on the object.

The analogical with  (51) relation in QSM, in contrast to one in
$\hbar kD$, has the form
\begin{equation}\label{55}
\frac{J}{S}\rightarrow\frac{\hbar\exp(-\hbar\omega/k_BT)}
{k_B(\hbar\omega/k_BT)\exp(-\hbar\omega/k_BT)}=\frac{T}{\omega}\rightarrow
0.
\end{equation}

 Therefore it is now possible to compare two theories ($\hbar kD$
 and QSM) experimentally by measuring the limiting value of this
 ratio.

 The first indication that the quantity $\varkappa$ plays an
 important role was obtained in Andronikashvili's experiments (1948) on
 the viscosity of liquid Helium below the $\lambda-$ point. There
 are
 also  another areas of Physics where the constant $\varkappa$
 appears.
 Now the constant $\varkappa$ is also observed in cold atomic gases and
 different solids, including graphene, as a characteristic of the     fundamentally   new state of
 matter - nearly perfect fluids (Shaefer, Teaney, 2009). We   hope that $\hbar kD$ can serve as an initial microtheory for constructing of modified thermodynamics fit for very small objects at ultra-low temperatures.

 This work was supported by the Russian Foundation for Basic
 Research (project No. 10-01-90408)

\pagebreak

\end{document}